\documentclass[twocolumn,showpacs,amsmath,floatfix,prb,aps]{revtex4}
\usepackage{graphicx}  
\usepackage{dcolumn}   
\usepackage{bm}        
\usepackage{amssymb}   
\usepackage[usenames]{color}
\usepackage{url}
\usepackage{hyperref}
\begin{document}

\title{Large magnetocrystalline anisotropy in tetragonally distorted Heuslers: a systematic study.}

\author{Y.-i. Matsushita$^{1,2}$}
\author{G. Madjarova$^{1,3}$}
\author{J. K. Dewhurst$^1$}
\author{S. Shallcross$^4$}
\author{C. Felser$^5$}
\author{S. Sharma$^{1,6}$}
\author{E. K. U. Gross$^1$}

\affiliation{1 Max-Planck Institute of Microstructure Physics, Weinberg 2, D-06120 Halle, Germany}
\affiliation{2 Department of Applied Physics, The University of Tokyo, Tokyo 113-8656, Japan}
\affiliation{3 Department of Physical Chemistry, Faculty of Chemistry and Pharmacy, Sofia University, 1126 Sofia, Bulgaria.}
\affiliation{4 Lehrstuhl f\"ur Theoretische Festk\"orperphysik, Staudtstr. 7-B2, 91058 Erlangen, Germany.}
\affiliation{5 Max Planck Institute for Chemical Physics of Solids, N\"othnitzer Strasse 40, 01187 Dresden, Germany}
\affiliation{6 Department of Physics, Indian Institute of Technology, Roorkee, 247667 Uttarkhand, India}

\date{\today}

\begin{abstract}

With a view to the design of hard magnets without rare earths we explore the possibility of large magnetocrystalline anisotropy energies in Heusler compounds that are unstable with respect to a tetragonal distortion. We consider the Heusler compounds Fe$_2$YZ with Y = (Ni, Co, Pt), and Co$_2$YZ with Y = (Ni, Fe, Pt) where, in both cases, Z = (Al, Ga, Ge, In, Sn). We find that for the Co$_2$NiZ, Co$_2$PtZ, and Fe$_2$PtZ families the cubic phase is always, at $T=0$, unstable with respect to a tetragonal distortion, while, in contrast, for the Fe$_2$NiZ and Fe$_2$CoZ families this is the case for only 2 compounds -- Fe$_2$CoGe and Fe$_2$CoSn. For all compounds in which a tetragonal distortion occurs we calculate the MAE finding remarkably large values for the Pt containing Heuslers, but also large values for a number of the other compounds (e.g. Co$_2$NiGa has an MAE of -2.11~MJ/m$^3$). The tendency to a tetragonal distortion we find to be strongly correlated with a high density of states at the Fermi level in the cubic phase. As a corollary to this fact we observe that upon doping compounds for which the cubic structure is stable such that the Fermi level enters a region of high DOS, a tetragonal distortion is induced and a correspondingly large value of the MAE is then observed.

\end{abstract}

\pacs{}
\maketitle

\section{Introduction}

Underpinning a diverse range of modern technologies, from computer hard drives to wind turbines, are hard magnets. These are magnets in which the local moments all preferentially align along a certain crystallographic direction, and may be characterized by the energy difference with an unfavorable spatial direction, known as the magnetocrystalline anisotropy energy (MAE). Evidently the local moments of such magnets are very stable and so make excellent permanent magnets, hence their central role in various technologies\cite{spintronics}. Current production of hard magnets relies on alloys of rare earth elements, in particular neodymium and dysprosium, e.g. the ``neodymium magnet'' Nd$_2$Fe$_{14}$B. Such alloys, due to the localized nature of the open shell $f$-electrons of the rare earths, possess a very large spin orbit coupling and, as a consequence, very high MAE values. However, as the rare earths are both costly and highly polluting to extract from ore there is a current focus on the design of hard magnets without rare earths\cite{McCallum, Coey1, Coey2, Kramer}. Magnetic materials having a low crystal symmetry evidently possess a natural spatial anisotropy, and this in turn can lead to very large values of the MAE. Such low symmetry magnets therefore offer a promising design route towards the next generation of hard magnets. Accurate calculation of the MAE requires sophisticated and computationally expensive first principles calculations, making difficult the kind of high throughput search that might be expected to yield interesting high MAE materials. In this paper we show that for a promising materials class - the Heusler alloys - the density of the states at the Fermi level provides a very good indicator of the propensity to distortion, and therefore of the likelihood of finding a high MAE material within this class. The use of such material markers for high MAE can, we believe, significantly alleviate the computation bottleneck preventing high throughput search.

The Heusler materials have attracted sustained attention due both to their exceptional magnetic properties as well as a huge variety of possible compounds that may be experimentally realized\cite{Lue, Alijani}; reviews may be found in Refs.~\onlinecite{Felser, Graf1, Graf2, Kreiner}. These materials, which consist of 4 inter-penetrating face centred cubic lattices, often exhibit a symmetry lowering structural transition to a tetragonal or hexagonal phase\cite{roy16,woll15,tal15,hong13,wint12}, raising the possibility of a crystal symmetry lowering induced large MAE. For Mn rich Heusler alloys this has previously been explored\cite{woll14,woll15}; here we consider this possibility in the Heusler families Fe$_2$YZ with Y = (Ni, Co, Pt), and Co$_2$YZ with Y = (Ni, Fe, Pt) where, in both cases, Z = (Al, Ga, Ge, In, Sn).

\begin{figure*}
\includegraphics[width=0.9\linewidth]{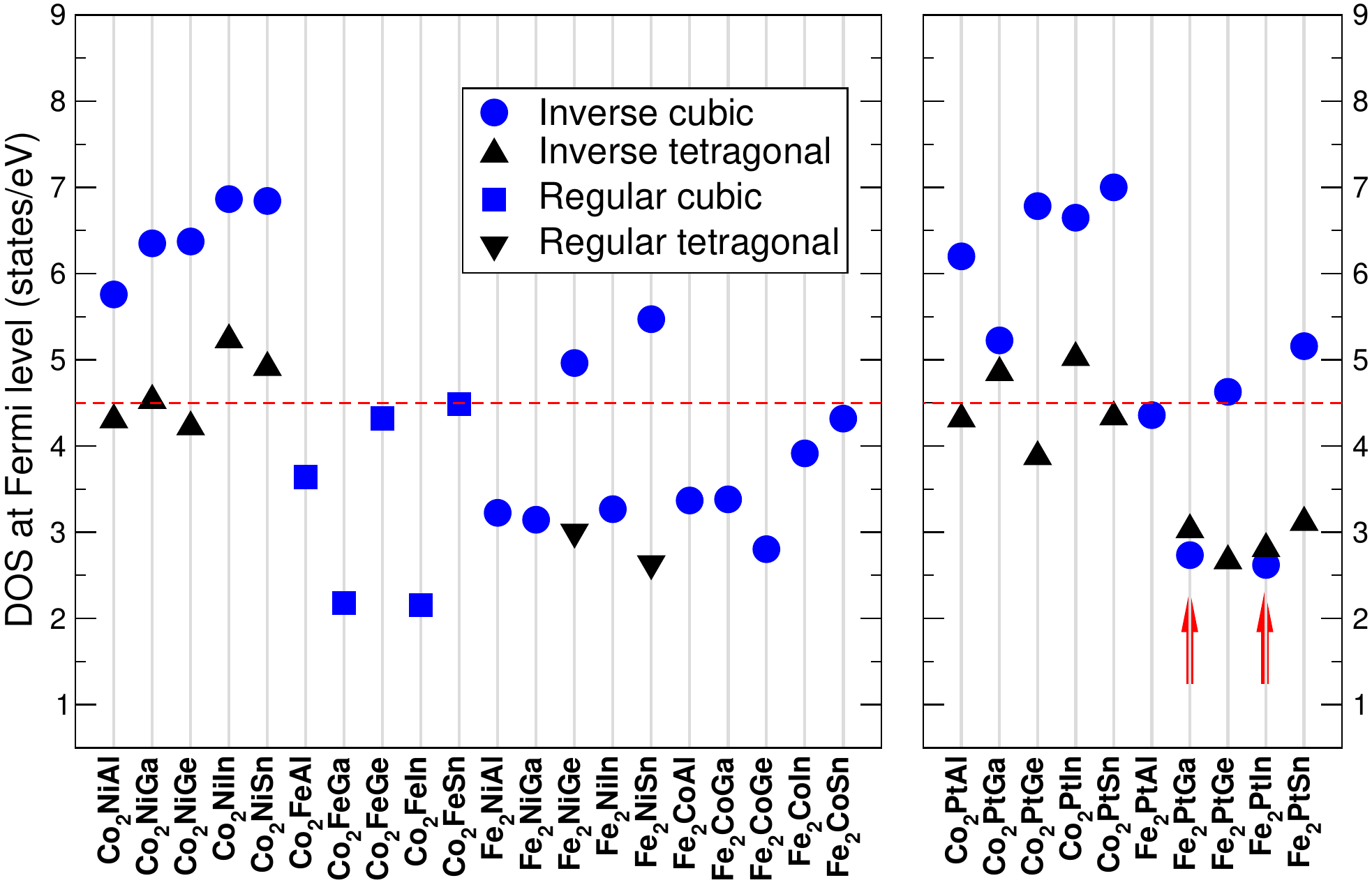}
\caption{Calculated total density of states at the Fermi energy for the Heusler compounds Co$_2$NiZ, Co$_2$FeZ, Co$_2$PtZ, Fe$_2$NiZ, Fe$_2$CoZ and Fe$_2$PtZ where Z = Al, Ge, Ga, Sn, or In. In each case the structure is indicated by the caption; and the tetragonal phase is shown when it is the lowest energy. Evidently, the instability of the cubic phase strongly correlates to the density of states at the Fermi energy. For almost all cases when DOS at Fermi level $>$ 4.5 states/eV (indicated by the dashed horizontal line) the cubic phase is unstable; two exceptions to this are indicated by arrows.}
\label{DOS}
\end{figure*}

Our principle findings are that (i) the Co$_2$NiZ and Co$_2$PtZ classes naturally distort to a tetragonal structure with $c/a$ values in the range 1.3-1.5; (ii) the Fe rich Heuslers generally do not distort, with the exceptions of Fe$_2$CoZ where Z = Ge or Sn and the Fe$_2$PtZ family; (iii) this distortion can induce a very high MAE - of up to 5 MJ/m$^3$ for the Pt containing Heuslers, comparable to the best known transition metal magnet L1$_0$-FePt, and of up to 1~MJ/m$^3$ for the Co rich but Pt free Heuslers. In each case where a distortion occurs the volume change is found to be very small (a few percent at most), with the exception of Fe$_2$PtGe in which a 6\% increase of volume occurs upon distortion. 

We furthermore find that this tendency to tetragonal distortion strongly correlates to a rather simple material descriptor, namely the density of states (DOS) at the Fermi level. A high DOS favours tetragonal distortion and, on this basis, we consider the possibility of inducing a tetragonal distortion by moderate doping (via a virtual crystal approximation) that shifts the Fermi energy from a low to a high DOS position. Consistent with the validity of this material descriptor  we find that the Heusler alloys Co$_2$FeAl and Co$_2$FeSn - in which the Fermi energy lies far from and close to a high DOS region respectively - all spontaneously suffer tetragonal distortion upon doping.

\section{Calculation details}

For structural relaxation we use the Vienna ab initio simulation package (VASP)\cite{vasp} with projector augmented wave (PAW) pseudopotentials \cite{PAW}, a plane-wave-basis set energy cutoff of 400 eV, and the Perdew-Burke-Ernzerhof (PBE) functional \cite{PBE}. Reciprocal space integration has been performed with a $\Gamma$-centered Monkhorst-Pack 10x10x10 mesh. The structural optimization has been converged to a tolerance of $10^{-5}$ eV, whereas the MAE values were obtained with a tolerance of $10^{-7}$ eV. All calculations are performed in the presence of spin-orbit coupling term. For the calculation of the MAE we have also deployed the all-electron full-potential linearized augmented-plane wave (FP-LAPW) code Elk\cite{elk}. The definition of MAE adopted in this study is the following:

\begin{eqnarray}
{\rm MAE}= E^{\rm tot}_{[100]}-E^{\rm tot}_{[001]},
\end{eqnarray}
where $E^{\rm tot}_{[100]}$ ($E^{\rm tot}_{[001]}$) represents the total energy with spin orientation in the [100] ([001]) direction. A positive value of the MAE therefore indicates that out-of-plane spin configuration is energetically favourable, whereas a negative one that the in-plane direction is favourable. 

The Heusler structure is described by the X$_2$YZ general formula, in which the species X and Y are transition metal elements whereas the Z atom is $p$-orbital element with metal character (from III or IV main groups). The crystal structure consists of four inter-penetrating face centred cubic lattices and belongs to the 225 (Fm-3m) symmetry group for the regular Heusler structure, and 216 (F-43m) for the inverse Heusler; Wyckoff positions of the atoms are presented in Table.~I.

\begin{table}[htb]
  \begin{tabular}{|c|c|c|c|c|}
  \hline
 & 4a & 4c &  4b & 4d\\
\hline
& (0,0,0) & (1/4,1/4,1/4) & (1/2,1/2,1/2) & (3/4,3/4,3/4)\\
\hline
Regular & Z & X & Y & X\\
\hline
Inverse & Z & Y & X & X\\
\hline
 \end{tabular}
  \label{tab}
  \caption{Structural order of regular and inverse Heusler structures}
\end{table} 

\section{Structural distortion}

\begin{figure}
\includegraphics[width=0.9\linewidth]{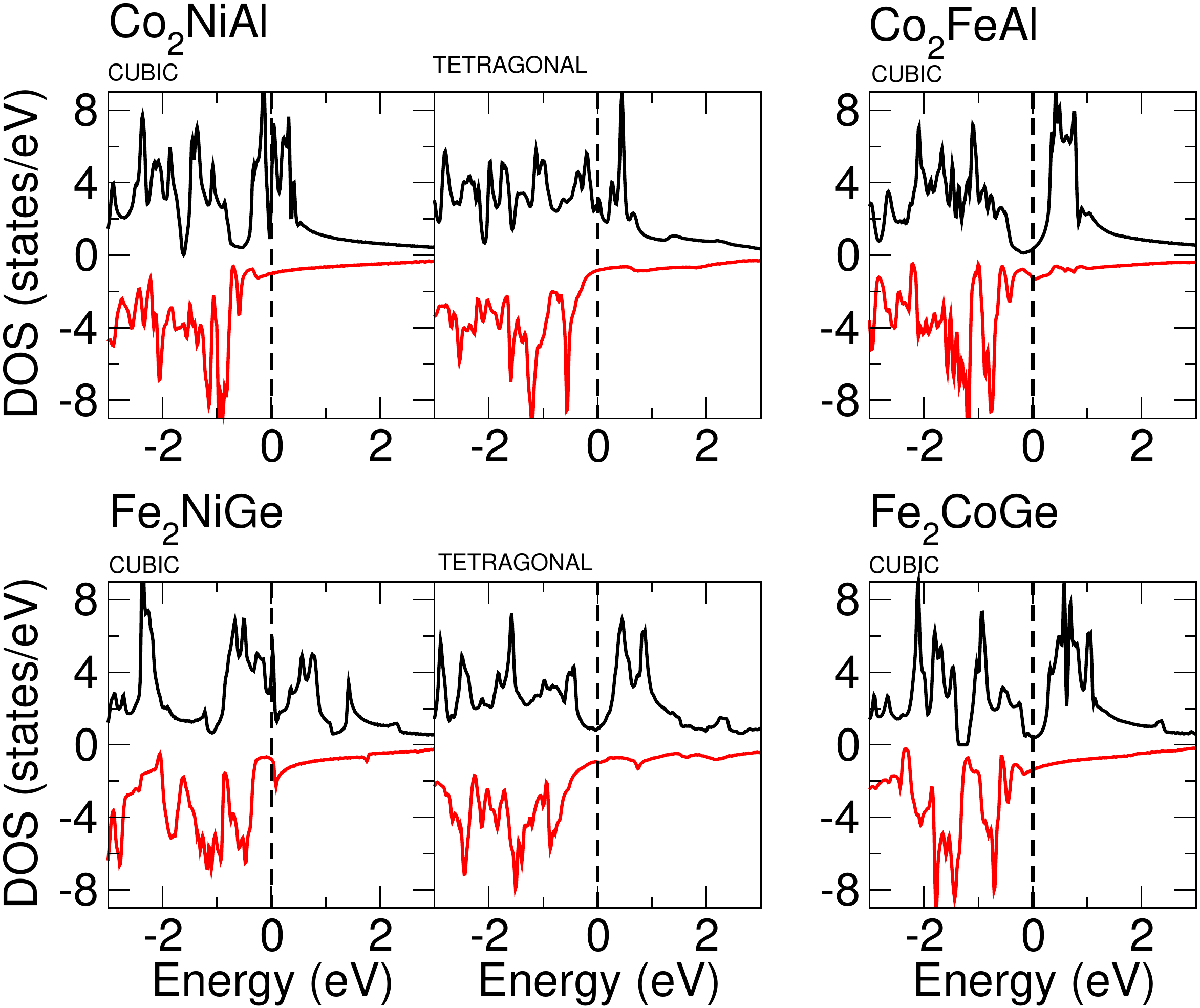}
\caption{Calculated total density of states for Co$_2$NiAl, Co$_2$FeAl, Fe$_2$NiGe and Fe$_2$CoGe in cubic and (where it is the lowest energy structure) the tetragonal phase. The Fermi energy is set to 0 and positive (negative) value of the DOS represents the minority (majority) spin projection.}
\label{TDOS}
\end{figure}

\begin{figure}
\includegraphics[width=0.9\linewidth]{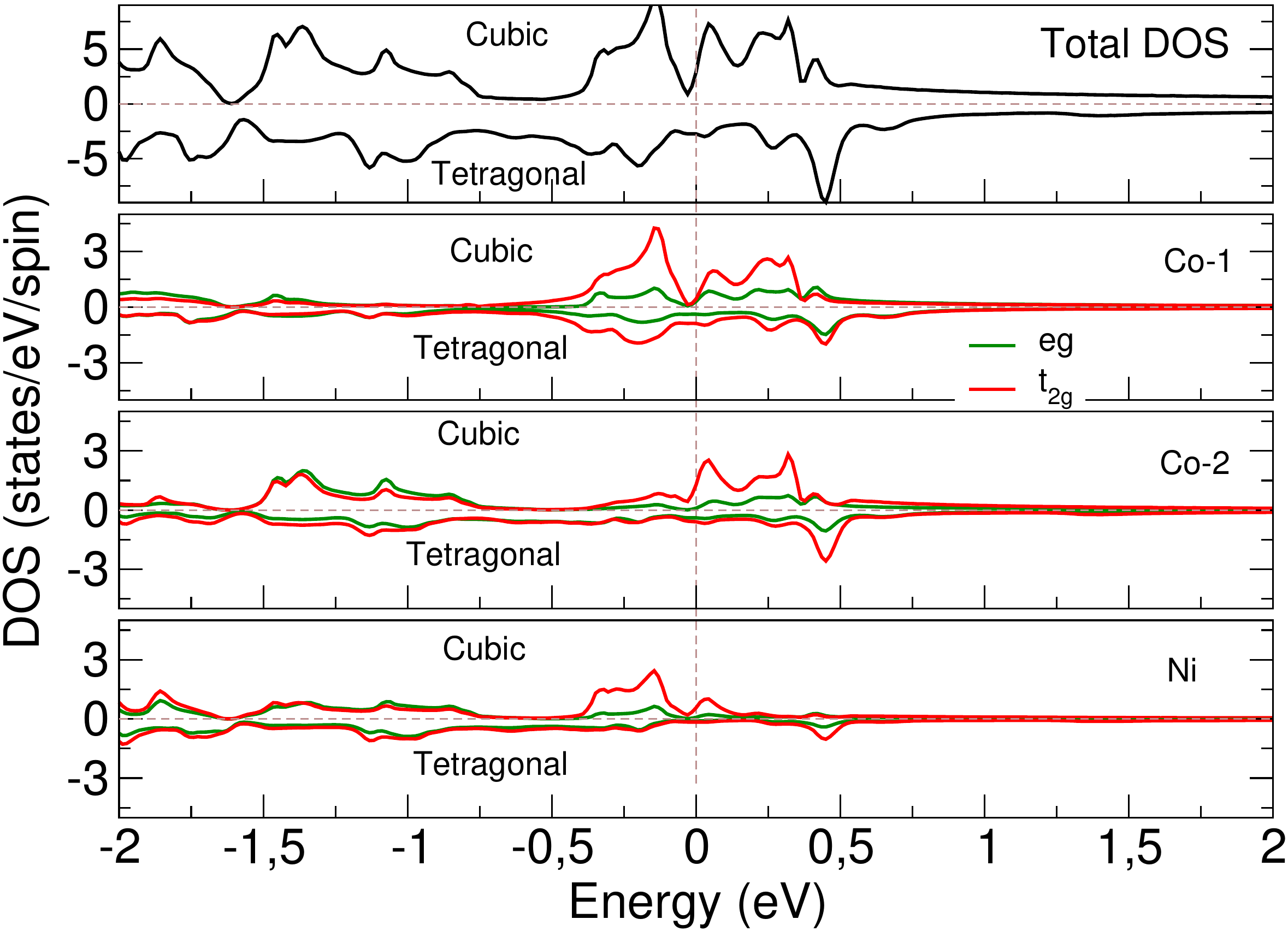}
\caption{Majority spin projected and as well as Co- $t_{2g}$ and $e_g$ projected DOS for Co$_2$NiGe; the cubic majority spin DOS is assigned a positive value, and the tetragonal majority spin DOS a negative value. The reduction in spectral weight near the Fermi energy that occurs due to the tetragonal distortion may clearly be seen. As may be seen from the lower panels, the redistribution of spectral weight upon tetragonal distortion consists primarily of (i) a reduction in the $e_g$ peak of Co-1 character and (ii) a reduction in the Co-2 $e_g$ peak at the Fermi level. A similar picture is found for other Co$_2$NiZ compounds as well as the Co$_2$PtZ compounds.}
\label{cdos}
\end{figure}

We first consider the stability with respect to tetragonal distortion of the Heusler alloys $X_2YZ$ in which the $X$ sub-lattices are occupied by either Fe or Co, the $Y$ sub-lattice by Fe, Co, Ni, or Pt, and $Z$ sub-lattice by Al, Ge, Ga, Sn, or In. This represents 30 materials in total, of which 10 have been previously experimentally synthesized; for details we refer the reader to Table I and II of Appendix A. In Fig.~\ref{DOS} we present the DOS at the Fermi energy of each of these Heusler materials for both the high symmetry cubic phase and, where it exists, the tetragonal structure. For the Co rich Heuslers Co$_2$NiZ and Co$_2$PtZ the high symmetry phase is always unstable with respect to tetragonal distortion while, in contrast, in the case of the Fe containing Heuslers the cubic phase is generally stable. There are two exceptions to this latter rule: Fe$_2$NiGe and Fe$_2$NiSn, and the Fe$_2$PtZ family. For all cases where the tetragonal phase is stable we find the $c/a$ ratios in the range 1.3-1.5 with the high end $c/a$ ratios found for the Fe$_2$YZ Heuslers in which Z is either Ge or Sn (curiously, as we will see, these are also the Heusler compounds that have the desired positive MAE). Of the Heuslers in Fig.~1 that have been experimentally synthesized only one, Co$_2$NiGa, is observed in the tetragonal structure, in agreement with our calculations; all others are found to be cubic, also in agreement with our calculations (with the exception of Fe$_2$NiGe for which we predict a tetragonal structure, this case will be discussed in detail below). 

\begin{figure*}
\includegraphics[width=0.9\linewidth]{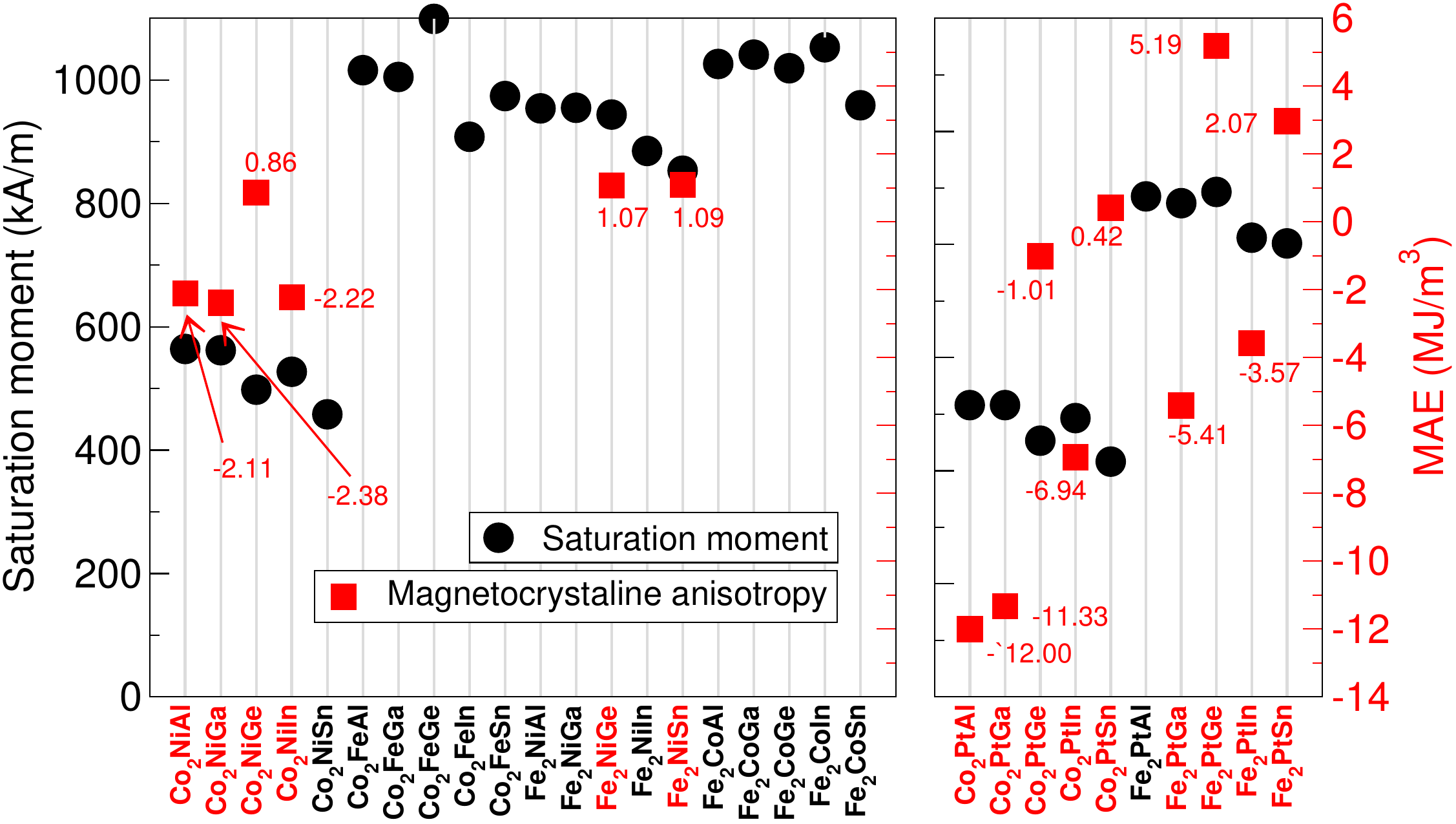}
\caption{Magnetic moments and MAE for the Heusler compounds Co$_2$NiZ, Co$_2$FeZ, Fe$_2$NiZ, and Fe$_2$CoZ where Z = Al, Ge, Ga, Sn, or In. Values of the MAE are presented only for tetragonally distorted Heusler compounds; the MAE for the cubic phase is, in comparison, negligibly small. A positive value of the MAE indicates an out-of-plane easy axis (i.e., the moments are aligned with the distortion axis), while a negative value an in-plane easy axis (i.e., the moments lie within the plane perpendicular to the distortion direction). For comparison, the value of MAE for Nd$_2$Fe$_{14}$B is 4.4 [MJ/m$^3$], and saturation magnetization is 1280 [kA/m].}
\label{MAG}
\end{figure*}

For each structure we have also determined whether the material takes on a regular or inverse occupation of the sub-lattices. As may be seen in Fig.~\ref{DOS} most of the structures are inverse Heusler except for the Co$_2$FeZ family where the regular cubic structure has a lower ground state energy. This finding is in a good agreement with an empirical rule first stated in Ref.~\onlinecite{Graf2}: when the electronegativity of the Y element is larger than that of the X element the system prefers the inverse Heusler structure, with otherwise the regular Heusler structure realized. There are, however, two deviations from this rule in our results. We find that Fe$_2$NiGe and Fe$_2$NiSn adopt a tetragonally distorted regular structure, in agreement with previous theoretical work\cite{gill10}, but in contrast to the inverse structure expected on the basis of the empirical rule (the electronegativity of Ni is higher than that of Fe).

In Ref.~\onlinecite{Gasi} experiment reports, in agreement with the semi-empirical rule, a cubic inverse structure for Fe$_2$NiGe. Accompanying theoretical calculations\cite{Gasi}, however, find that the energy change due to antisite disorder is always much smaller than the thermal energy available due to annealing (which takes place at 650~K in the experiment). The authors of Ref.~\onlinecite{Gasi} therefore conclude that annealing will control the state of order for the Fe$_2$NiZ family. The mismatch between experiment and our results, calculated for fully ordered structures, therefore likely has its origin in thermal induced substitutional disorder. It is worth pointing out that the energy difference between the tetragonally distorted regular structure (our lowest energy structure) and the inverse cubic structure is 120~meV, i.e. about double the thermal energy due to annealing. This indicates that the presence of antisite disorder has a significant impact on the propensity of this material towards tetragonal disorder and that, at least for the Fe$_2$NiZ family, the coupling between disorder and tetragonal distortion is a subject worthy of further investigation.

We now consider the electronic origin of the instability of the cubic phase with respect to a tetragonal distortion. Such instability of the high symmetry phase has been observed in many Heusler compounds, in particular the Mn rich Heuslers\cite{book,woll14,woll15}, and has been attributed to a number of different mechanisms: a Jahn-Teller effect\cite{book}, a ``band'' JT effect\cite{Rh}, a nesting induced Fermi surface instability\cite{bar07}, and anomalous phonon modes\cite{zay05,paul15}.

In Fig.~\ref{TDOS} we present the total density of states for four representative examples of the set of Heusler compounds we investigate. For all four cases (and for all Heuslers we study in this work) the minority spin channel is not significantly involved in the mechanism of distortion, having a very low DOS near the Fermi energy. For the cases (Co$_2$NiAl, Fe$_2$NiGe) in which the cubic phase is unstable we see a clear redistribution of spectral weight near the Fermi energy, such that a high DOS near $E_F$ is lowered by the opening up of a  ``valley'' near $E_F$ in the tetragonal phase. On the other hand, for the materials in which the cubic phase is stable the DOS at $E_F$ is already very low (see the right hand panels of Fig.~\ref{DOS} for the representative cases of Co$_2$FeAl and Fe$_2$CoGe). As may be seen in Fig.~\ref{cdos} for the case of Co$_2$NiAl this redistribution of weight occurs in all species and momentum channels, but with states of Co character being the more important. Interestingly, it is seen that the redistribution occurs particularly in states of $e_g$ character.

\section{Magnetic moments and magnetocrystalline anisotropy}

In Fig.~\ref{MAG} we present the total magnetic moment, saturation magnetization M$_s$ and MAE for the Co$_2$YZ and Fe$_2$YZ Heusler families. In all systems the magnetic order is found to be ferromagnetic. To a good approximation the values of the saturation magnetization M$_s$ fall into four distinct bands: (i) $M_s$ close to 500~kA/m for Co$_2$NiZ; (ii) $M_s$ close to 900~kA/m for Co$_2$FeZ, Fe$_2$NiZ, and Fe$_2$CoZ; (iii) $M_s$ close to 500~kA/m for Co$_2$PtZ; and (iv) $M_s$ close to 800~kA/m for Fe$_2$PtZ. From the viewpoint of hard magnetic applications a high value of the saturation magnetization is desired, and from this point of view the Co$_2$FeZ, Fe$_2$NiZ, Fe$_2$CoZ, and Fe$_2$PtZ compounds are most interesting. For comparison we recall that the two ``standard'' hard magnets have saturation magnetizations of 970~kA/m for SmCo$_5$ and 1280~kA/m for Nd$_2$Fe$_{14}$B.

\begin{figure}
\includegraphics[width=0.9\linewidth]{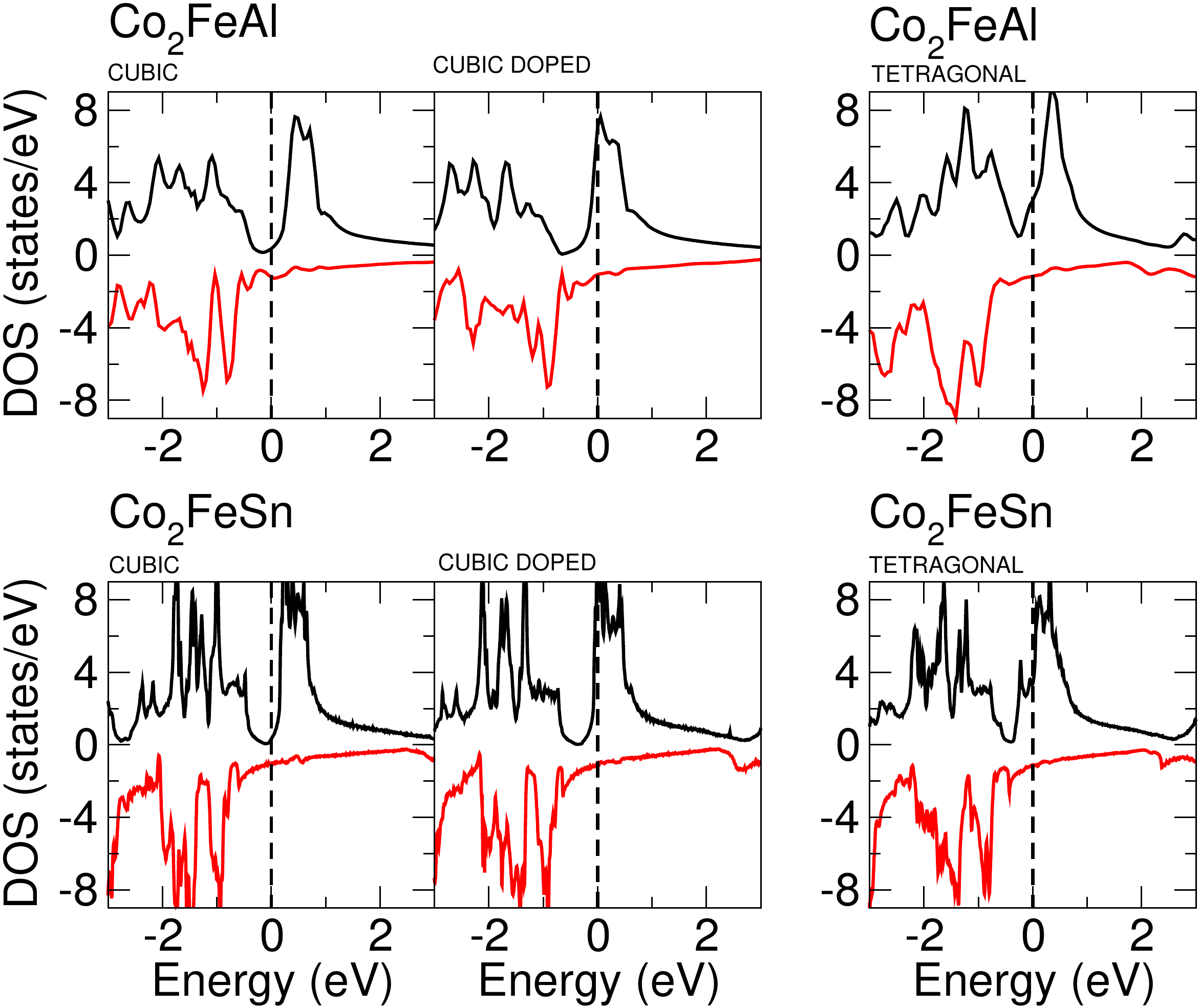}
\caption{Calculated density of states of Co$_2$FeAl and Co$_2$FeSn for the cubic phase, the cubic phase with doping of 1.5e and 0.3e respectively, and the tetragonally distorted phase. In each case the electron doping shifts the Fermi energy to a region of high density of states and drives a tetragonal distortion of the cubic phase, which is otherwise stable. Positive (negative) value of DOS represents the minority (majority) spin projection.}
\label{DOS_doped}
\end{figure}

We now turn to a discussion of the MAE values realized in the cases for which a tetragonal distortion occurs (see also Fig.~\ref{MAG}). We first note that a positive value of the MAE indicates that the magnetic moments all align with the symmetry axis of the tetragonally distorted material: this is essential for hard magnetic applications. When the MAE takes on a negative value this indicates that the moments are in the plane perpendicular to this symmetry axis. We have checked the energy required to rotate spins in-plane finding, as expected, a very soft energy dependence. This freedom to rotate the spin structure obviously renders such cases entirely unsuitable for hard magnetic application. We will therefore focus on those cases for which the MAE is positive.

\begin{table*}[htb]
  \begin{tabular}{|l|c|c|c|c|}
  \hline
    & structure &~ $\mu_{\rm B}^{\rm tot}~$ &a$^{\rm calc}$, c$^{\rm calc}$~[\AA]& MAE [MJ/m$^3$]\\ \hline
    ~~Co$_2$FeAl~~& regular cubic &5.08  & a=c=5.69& $-$\\ \hline
    ~~1.5e$^-$doped-Co$_2$FeAl~~& regular tetragonal &5.43  & a=6.16 c=6.88&$-$0.94 \\ \hline
    ~~Co$_2$FeSn~~& regular cubic & 5.66 & a=c=5.64& $-$  \\ \hline
    ~~0.3e$^-$doped-Co$_2$FeAl~~& regular tetragonal &5.41  & a=5.97 c=6.45& $-$1.30\\ \hline
  \end{tabular}
  \label{table_doped}
  \caption{Calculated material properties of Co$_2$FeAl, Co$_2$FeSn, and their electron doped systems. The most stable structure is shown in the 2nd column. The calculated magnetic moments per formula unit, lattice parameters, $a$ and $c$, and the magnetocrystalline anisotropy energies for the tetragonal cases are also listed.}
\end{table*}

Of the 15 compounds that suffer a tetragonal distortion only 6 have $E_{MAE} > 0$. Curiously, these are the compounds for which the Z element is either Ge or Sn: Co$_2$NiGe, Fe$_2$NiGe, Fe$_2$NiSn, Co$_2$PtSn, Fe$_2$PtGe, and Fe$_2$PtSn. The values of the MAE for the Pt free compounds are all, as expected, modest as compared to the Pt containing compounds. The maximum positive MAE for Pt free compounds are found in Fe$_2$NiGe and Fe$_2$NiSn, with an MAE of $\approx 1$~MJ/m$^3$, while for the Pt containing compounds we find a much higher MAE of 5.19~MJ/m$^3$ for Fe$_2$PtGe. This value is close to the currently highest value observed for an MAE in a rare earth free material (a value of 7~MJ/m$^3$ for L1$_0$-FePt\cite{FePt_expt}). The rather high M$_s$ value of 516~kA/m suggests this material might be interesting to further explore in the context of specialist application as a hard magnet.

\section{Distortion control}

The previous two sections lead us to conclude that (i) the propensity to tetragonal distortion strongly correlates with a high DOS at the Fermi energy in the cubic phase and (ii) that if a tetragonal distortion occurs, high values of the MAE are possible. This raises the possibility of, with a view to engineering a high MAE, inducing such a distortion by doping.

To this end we consider the two materials presented in Fig.~\ref{TDOS} in which the Fermi energy lies in the valley between the two high DOS regions, and dope the cubic phase within the virtual crystal approximation (VCA). In the case of Co$_2$FeAl a doping of 1.5 electrons is required to shift the Fermi energy into the high DOS region, with a more modest 0.3 electrons required in the case of Fe$_2$CoGe. In both cases we find that upon such doping, the cubic phase becomes unstable with respect to a tetragonal distortion; structural details may be found in Table~II. 
A subsequent calculation of the MAE finds values comparable to those obtained for the naturally tetragonally distorting Heusler compounds. It is also interesting to note that the mechanism of the distortion appears to be somewhat different from the ``natural'' cases: while in Fig.~\ref{TDOS} it is clearly seen that the distortion results in a significant redistribution of spectral weight away from the Fermi energy via the opening of a ``repulsion valley'' in Fig.~\ref{DOS_doped} this effect is seen to be much weaker. This of course, may be an artifact of the VCA.

\section{Conclusion}

We have addressed the question of whether we may obtain large magnetocrystalline anisotropy energies in Heusler compounds that adopt a low symmetry tetragonal structure. To this end we have investigated the Heusler compounds Fe$_2$YZ with Y = (Ni, Co, Pt), and Co$_2$YZ with Y = (Ni, Fe, Pt) where, in both cases, Z = (Al, Ga, Ge, In, Sn). We find that the cubic phase of 15 of these 30 Heusler compounds is unstable with respect to tetragonal distortion, in particular for the Co$_2$NiZ, Co$_2$PtZ, and Fe$_2$PtZ families the cubic phase is always, at $T=0$, unstable. In contrast, for the Fe$_2$NiZ and Fe$_2$CoZ families this is the case for only 2 compounds -- Fe$_2$CoGe and Fe$_2$CoSn. The mechanism behind this distortion involves a significant redistribution of spectral weight near the Fermi energy, such that a ``valley'' in the DOS at the Fermi energy is opened up in the tetragonal phase leading to a reduction in the number of states near the Fermi energy. Curiously, we find that for the compounds we investigate a good rule of thumb exists that if the DOS at the Fermi level is greater than 4.5~states/eV, the cubic phase is unstable.

Of the 15 compounds that suffer tetragonal distortion the magnetocrystalline anisotropy energies are found to range in values from -12~MJ/m$^3$ (Co$_2$PtAl) to +5.19~MJ/m$^3$ (Fe$_2$PtGe). As expected, the values of the MAE for the Pt free Heuslers are more modest in magnitude, and range in value from -2.38~MJ/m$^3$ (Co$_2$NiGa) to 1.09~MJ/m$^3$ (Fe$_2$NiSn). For hard magnet application only positive values of the magnetocrystalline anisotropy energies, which correspond to moments aligned with the tetragonal symmetry axis, are interesting. Interestingly, we find that the MAE takes on a positive value for all cases in which the Z element is either Ge or Sn.

Finally, we have considered the possibility of doping the Heusler compounds in which the cubic phase is stable in order to induce a tetragonal distortion. Using the virtual crystal approximation we find that this is indeed possible, and the doping induced distortion results in magnetocrystalline anisotropy energies values comparable to those obtained in the naturally distorting Heusler compounds.

\section{Acknowledgments}
YM, GM and S. Sharma would like to thank the Heusler project funded by the MPG. 

\begin{appendix}
\section{Details of the structural and magnetic properties of the Heuslers investigated in this work}
\label{appA}

In this Appendix we present structural details of the Heusler compounds calculated in the manuscript along with experimental structure data where this exists. In Table~III 
we present the Heusler compounds Co$_2$NiZ, Co$_2$FeZ, and in Table~IV 
the compounds Fe$_2$NiZ, and Fe$_2$CoZ where in each case Z = Al, Ge, Ga, Sn, or In.

\begin{table*}[htb]
  \begin{tabular}{|l|c|c|c|c|c|c|}
  \hline
    & structure &~ $ Moment(\mu_{\rm B}) ~$ & a$^{\rm calc}$, c$^{\rm calc}$~[\AA] &M$_{\rm s}$ [kA/m]& MAE [MJ/m$^3$]&~ {\rm expt.}\\ \hline
    ~~Co$_2$NiAl~~& inv.~tet. & 2.78 (Co=1.4, 1.2 Ni=0.2) & a=5.20 c=6.76&  564 & $-$2.11 & \\ \hline
    ~~Co$_2$NiGa~~& inv.~tet. & 2.79 (Co=1.6, 1.4 Ni=0.2) & a=5.19 c=6.80&  562 & $-$2.38 &$\mu_{\rm B}^{\rm tot}$=2.807, a=3.669 c=7.331\cite{Co2NiGa}\\ \hline
    ~~Co$_2$NiGe~~& inv.~tet. & 2.46 (Co=1.3, 1.0 Ni=0.2) & a=5.12 c=6.97& 498 & 0.86& \\ \hline
    ~~Co$_2$NiIn~~& inv.~tet. & 2.99 (Co=1.6, 1.3 Ni=0.2) & a=5.43 c=7.13&527 & $-$2.22& \\ \hline
    ~~Co$_2$NiSn~~& inv.~tet. & 2.58 (Co=1.4, 1.1 Ni=0.2) & a=5.37 c=7.22& 458 & 0 &   \\ \hline
    ~~Co$_2$FeAl~~& reg.~cubic & 5.08 (Co=1.2, 1.2 Fe=2.8)& a=c=5.69 &  1016 & $-$&$\mu_{\rm B}^{\rm tot}$=4.82-5.22, a=c=5.74 \cite{Co2FeAl1, Co2FeAl2}\\ \hline
    ~~Co$_2$FeGa~~& reg.~cubic & 5.07 (Co=1.2, 1.2 Fe=2.8)& a=c=5.70 &  1005 & $-$&$\mu_{\rm B}^{\rm tot}$=5.035, a=c=5.727 \cite{Co2FeGa} \\ \hline
    ~~Co$_2$FeGe~~& reg.~cubic & 5.60 (Co=1.4, 1,4 Fe=2.9) & a=c=5.74 & 1099 & $-$&$\mu_{\rm B}^{\rm tot}$=5.54-5.70, a=c=5.702 \cite{Co2FeGe} \\ \hline
    ~~Co$_2$FeIn~~& reg.~cubic & 5.23 (Co=1.3, 1.3 Fe=2.8)& a=c=5.64 & 908 & $-$&   \\ \hline
    ~~Co$_2$FeSn~~& reg.~cubic & 5.66 (Co=1.4, 1.4 Fe=2.9)& a=c=5.64 & 974 & $-$&    \\ \hline
    ~~Co$_2$PtAl~~& inv.~tet. & 2.84  (Co=1.5, 1.2 Pt=0.1)  & a=5.36 c=7.13& 516 & $-$12 &\\ \hline
    ~~Co$_2$PtGa~~& inv.~tet. & 2.88 (Co=1.6, 1.3 Pt=0.1)  & a=5.36 c=7.21& 516 & $-$11.33 &\\ \hline
    ~~Co$_2$PtGe~~& inv.~tet. & 2.52 (Co=1.5, 1.0 Pt=0.1)  & a=5.32 c=7.31& 453 & $-$1.01& \\ \hline
    ~~Co$_2$PtIn~~& inv.~tet. & 3.08 (Co=1.6, 1.4 Pt=0.1)  & a=5.50 c=7.67& 493 & $-$6.94 &  \\ \hline
    ~~Co$_2$PtSn~~& inv.~tet. & 2.59 (Co=1.5, 1.1 Pt=0.1)  & a=5.50 c=7.63& 416 & 0.42  &   \\ \hline
  \end{tabular}
    \label{TCo}
  \caption{Calculated material properties of Co-family Heusler materials in cubic and tetragonal structures. The most stable structure in regular Heusler (denoted by "reg.") and inverse Heusler (denoted by "inv.") with cubic or tetragonal (denoted by "tet.") symmetry are shown in the 2nd column. The calculated total energy and magnetic moments per formula unit, distortion parameter c/a, saturated magnetic moment M$_{\rm s}$, atom resolved moments, and MAE values are also listed. Note that two X atoms are not equivalent in inverse Heusler with cubic or tetragonal symmetry. Therefore, two values of atom resolved moments for X elements are listed in inverse Heusler materials.}
\end{table*}

\begin{table*}[htb]
  \begin{tabular}{|l|c|c|c|c|c|c|}
  \hline
    & structure &~ $ Moment(\mu_{\rm B}) ~$ &a$^{\rm calc}$, c$^{\rm calc}$~[\AA]& M$_{\rm s}$ [kA/m]& MAE [MJ/m$^3$]&~ {\rm expt.}\\ \hline
    ~~Fe$_2$NiAl~~& inv.~cubic & 4.86 (Fe=2.6, 1.8 Ni=0.5)& a=c=5.74&  954 & $-$&$\mu_{\rm B}^{\rm tot}$=4.00-4.46, a=c=5.75 \cite{Zhang, Fe2NiAl2, Fe2NiAl3}\\ \hline
    ~~Fe$_2$NiGa~~& inv.~cubic & 4.91 (Fe=2.6, 1.9 Ni=0.4)& a=c=5.77 &  955 & $-$&$\mu_{\rm B}^{\rm tot}$=4.29-4.89, a=c=5.80\cite{Zhang, Gasi} \\ \hline
    ~~Fe$_2$NiGe~~& reg.~tet. & 4.86  (Fe=2.3, 2.3 Ni=0.3)& a=5.02 c=7.59 & 944 & 1.07&$\mu_{\rm B}^{\rm tot}$=4.20-4.38, a=c=5.76 \cite{Zhang, Gasi} \\ \hline
    ~~Fe$_2$NiIn~~& inv.~cubic & 5.22 (Fe=2.7, 2.2 Ni=0.4)& a=c=6.07 & 885 & $-$ &  \\ \hline
    ~~Fe$_2$NiSn~~& reg.~tet. & 5.02 (Fe=2.4, 2.4 Ni=0.3)& a=5.22 c=8.02 & 853 & 1.09&  \\ \hline
    ~~Fe$_2$CoAl~~& inv.~cubic & 5.14 (Fe=2.5, 1.6 Co=1.0)& a=c=5.71 &  1026 & $-$&$\mu_{\rm B}^{\rm tot}$=4.91, a=c=5.766 \cite{Fe2NiAl3}\\ \hline
    ~~Fe$_2$CoGa~~& inv.~cubic & 5.28 (Fe=2.5, 1.8 Co=1.1)& a=c=5.76 &  1041 & $-$&$\mu_{\rm B}^{\rm tot}$=5.09, a=c=5.767 \cite{Fe2NiAl2}\\ \hline
    ~~Fe$_2$CoGe~~& inv.~cubic & 5.13 (Fe=2.7, 1.5 Co=1.0)& a=c=5.72 & 1019 & $-$&$\mu_{\rm B}^{\rm tot}$=5.06, a=c=5.775-5.764 \cite{Fe2NiGe} \\ \hline
    ~~Fe$_2$CoIn~~& inv.~cubic & 6.18 (Fe=2.7, 2.3 Co=1.3)& a=c=6.03 & 1053 & $-$&   \\ \hline
    ~~Fe$_2$CoSn~~& inv.~cubic & 5.54 (Fe=2.7, 1.9 Co=1.1)& a=c=5.99 & 959 & $-$ &   \\ \hline
    ~~Fe$_2$PtAl~~& inv.~cubic & 5.04 (Fe=2.8, 2.1 Pt=0.2)& a=c=5.97 &  885 & $-$& \\ \hline
    ~~Fe$_2$PtGa~~& inv.~tet. & 5.06 (Fe=2.7, 2.3 Pt=0.2)& a=5.50 c=7.10 &  873 & $-$5.41 & \\ \hline
    ~~Fe$_2$PtGe~~& reg.~tet. & 5.27 (Fe=2.6, 2.6 Pt=0.1)& a=5.40 c=7.38 & 893 & 5.19 & \\ \hline
    ~~Fe$_2$PtIn~~& inv.~tet. & 5.27 (Fe=2.8, 2.5 Pt=0.1)& a=5.64 c=7.56 & 812 & $-$3.57 &   \\ \hline
    ~~Fe$_2$PtSn~~& reg.~tet. & 5.26 (Fe=2.6, 2.6 Pt=0.1)& a=5.64 c=7.58& 802 & 2.97 &   \\ \hline
  \end{tabular}
  \label{TFe}
  \caption{Calculated material properties of Fe-family Heusler materials in cubic and tetragonal structures. The most stable structure in regular Heusler (denoted by "reg.") and inverse Heusler (denoted by "inv.") with cubic or tetragonal (denoted by "tet.") symmetry are shown in the 2nd column. The calculated total energy and magnetic moments per formula unit, distortion parameter c/a, saturated magnetic moment M$_{\rm s}$, atom resolved moments, and MAE values are also listed. Note that two X atoms are not equivalent in inverse Heusler with cubic or tetragonal symmetry. Therefore, two values of atom resolved moments for X elements are listed in inverse Heusler materials.}
\end{table*}

\end{appendix}


\begin{thebibliography}{38}%
\makeatletter
\providecommand \@ifxundefined [1]{%
 \@ifx{#1\undefined}
}%
\providecommand \@ifnum [1]{%
 \ifnum #1\expandafter \@firstoftwo
 \else \expandafter \@secondoftwo
 \fi
}%
\providecommand \@ifx [1]{%
 \ifx #1\expandafter \@firstoftwo
 \else \expandafter \@secondoftwo
 \fi
}%
\providecommand \natexlab [1]{#1}%
\providecommand \enquote  [1]{``#1''}%
\providecommand \bibnamefont  [1]{#1}%
\providecommand \bibfnamefont [1]{#1}%
\providecommand \citenamefont [1]{#1}%
\providecommand \href@noop [0]{\@secondoftwo}%
\providecommand \href [0]{\begingroup \@sanitize@url \@href}%
\providecommand \@href[1]{\@@startlink{#1}\@@href}%
\providecommand \@@href[1]{\endgroup#1\@@endlink}%
\providecommand \@sanitize@url [0]{\catcode `\\12\catcode `\$12\catcode
  `\&12\catcode `\#12\catcode `\^12\catcode `\_12\catcode `\%12\relax}%
\providecommand \@@startlink[1]{}%
\providecommand \@@endlink[0]{}%
\providecommand \url  [0]{\begingroup\@sanitize@url \@url }%
\providecommand \@url [1]{\endgroup\@href {#1}{\urlprefix }}%
\providecommand \urlprefix  [0]{URL }%
\providecommand \Eprint [0]{\href }%
\providecommand \doibase [0]{http://dx.doi.org/}%
\providecommand \selectlanguage [0]{\@gobble}%
\providecommand \bibinfo  [0]{\@secondoftwo}%
\providecommand \bibfield  [0]{\@secondoftwo}%
\providecommand \translation [1]{[#1]}%
\providecommand \BibitemOpen [0]{}%
\providecommand \bibitemStop [0]{}%
\providecommand \bibitemNoStop [0]{.\EOS\space}%
\providecommand \EOS [0]{\spacefactor3000\relax}%
\providecommand \BibitemShut  [1]{\csname bibitem#1\endcsname}%
\let\auto@bib@innerbib\@empty
\bibitem [{\citenamefont {Winterlik}\ \emph
  {et~al.}(2012{\natexlab{a}})\citenamefont {Winterlik}, \citenamefont
  {Chadov}, \citenamefont {Gupta}, \citenamefont {Alijani}, \citenamefont
  {Gasi}, \citenamefont {Filsinger}, \citenamefont {Balke}, \citenamefont
  {Fecher}, \citenamefont {Jenkins}, \citenamefont {Casper}, \citenamefont
  {Kubler}, \citenamefont {Liu}, \citenamefont {Gao}, \citenamefont {Parkin},\
  and\ \citenamefont {Felser}}]{spintronics}%
  \BibitemOpen
  \bibfield  {author} {\bibinfo {author} {\bibfnamefont {J.}~\bibnamefont
  {Winterlik}}, \bibinfo {author} {\bibfnamefont {S.}~\bibnamefont {Chadov}},
  \bibinfo {author} {\bibfnamefont {A.}~\bibnamefont {Gupta}}, \bibinfo
  {author} {\bibfnamefont {V.}~\bibnamefont {Alijani}}, \bibinfo {author}
  {\bibfnamefont {T.}~\bibnamefont {Gasi}}, \bibinfo {author} {\bibfnamefont
  {K.}~\bibnamefont {Filsinger}}, \bibinfo {author} {\bibfnamefont
  {B.}~\bibnamefont {Balke}}, \bibinfo {author} {\bibfnamefont
  {G.}~\bibnamefont {Fecher}}, \bibinfo {author} {\bibfnamefont
  {C.}~\bibnamefont {Jenkins}}, \bibinfo {author} {\bibfnamefont
  {F.}~\bibnamefont {Casper}}, \bibinfo {author} {\bibfnamefont
  {J.}~\bibnamefont {Kubler}}, \bibinfo {author} {\bibfnamefont {G.-D.}\
  \bibnamefont {Liu}}, \bibinfo {author} {\bibfnamefont {L.}~\bibnamefont
  {Gao}}, \bibinfo {author} {\bibfnamefont {S.}~\bibnamefont {Parkin}}, \ and\
  \bibinfo {author} {\bibfnamefont {C.}~\bibnamefont {Felser}},\ }\href@noop {}
  {\bibfield  {journal} {\bibinfo  {journal} {Adv. Mater.}\ }\textbf {\bibinfo
  {volume} {24}},\ \bibinfo {pages} {6283} (\bibinfo {year}
  {2012}{\natexlab{a}})}\BibitemShut {NoStop}%
\bibitem [{\citenamefont {McCallum}\ \emph {et~al.}(2014)\citenamefont
  {McCallum}, \citenamefont {Lewis}, \citenamefont {Skomski}, \citenamefont
  {Kramer},\ and\ \citenamefont {Anderson}}]{McCallum}%
  \BibitemOpen
  \bibfield  {author} {\bibinfo {author} {\bibfnamefont {R.~W.}\ \bibnamefont
  {McCallum}}, \bibinfo {author} {\bibfnamefont {L.~H.}\ \bibnamefont {Lewis}},
  \bibinfo {author} {\bibfnamefont {R.}~\bibnamefont {Skomski}}, \bibinfo
  {author} {\bibfnamefont {M.~J.}\ \bibnamefont {Kramer}}, \ and\ \bibinfo
  {author} {\bibfnamefont {I.~E.}\ \bibnamefont {Anderson}},\ }\href@noop {}
  {\bibfield  {journal} {\bibinfo  {journal} {Annu. Rev. Mater. Res.}\ }\textbf
  {\bibinfo {volume} {44}},\ \bibinfo {pages} {451} (\bibinfo {year}
  {2014})}\BibitemShut {NoStop}%
\bibitem [{\citenamefont {Coey}(2011)}]{Coey1}%
  \BibitemOpen
  \bibfield  {author} {\bibinfo {author} {\bibfnamefont {J.~M.~D.}\
  \bibnamefont {Coey}},\ }\href@noop {} {\bibfield  {journal} {\bibinfo
  {journal} {IEEE transactions on magnetics}\ }\textbf {\bibinfo {volume}
  {47}},\ \bibinfo {pages} {4671} (\bibinfo {year} {2011})}\BibitemShut
  {NoStop}%
\bibitem [{\citenamefont {Coey}(2012)}]{Coey2}%
  \BibitemOpen
  \bibfield  {author} {\bibinfo {author} {\bibfnamefont {J.~M.~D.}\
  \bibnamefont {Coey}},\ }\href@noop {} {\bibfield  {journal} {\bibinfo
  {journal} {Scripta Materialia}\ }\textbf {\bibinfo {volume} {67}},\ \bibinfo
  {pages} {524} (\bibinfo {year} {2012})}\BibitemShut {NoStop}%
\bibitem [{\citenamefont {Kramer}\ \emph {et~al.}(2012)\citenamefont {Kramer},
  \citenamefont {W.}, \citenamefont {I.A.},\ and\ \citenamefont
  {Constantinides}}]{Kramer}%
  \BibitemOpen
  \bibfield  {author} {\bibinfo {author} {\bibfnamefont {M.~J.}\ \bibnamefont
  {Kramer}}, \bibinfo {author} {\bibfnamefont {M.~R.}\ \bibnamefont {W.}},
  \bibinfo {author} {\bibfnamefont {A.}~\bibnamefont {I.A.}}, \ and\ \bibinfo
  {author} {\bibfnamefont {S.}~\bibnamefont {Constantinides}},\ }\href@noop {}
  {\bibfield  {journal} {\bibinfo  {journal} {JOM}\ }\textbf {\bibinfo {volume}
  {64}},\ \bibinfo {pages} {752} (\bibinfo {year} {2012})}\BibitemShut
  {NoStop}%
\bibitem [{\citenamefont {Lue}\ and\ \citenamefont {Kuo}(2002)}]{Lue}%
  \BibitemOpen
  \bibfield  {author} {\bibinfo {author} {\bibfnamefont {C.~S.}\ \bibnamefont
  {Lue}}\ and\ \bibinfo {author} {\bibfnamefont {Y.-K.}\ \bibnamefont {Kuo}},\
  }\href {\doibase 10.1103/PhysRevB.66.085121} {\bibfield  {journal} {\bibinfo
  {journal} {Phys. Rev. B}\ }\textbf {\bibinfo {volume} {66}},\ \bibinfo
  {pages} {085121} (\bibinfo {year} {2002})}\BibitemShut {NoStop}%
\bibitem [{\citenamefont {Alijani}\ \emph {et~al.}(2011)\citenamefont
  {Alijani}, \citenamefont {Ouardi}, \citenamefont {Fecher}, \citenamefont
  {Winterlik}, \citenamefont {Naghavi}, \citenamefont {Kozina}, \citenamefont
  {Stryganyuk}, \citenamefont {Felser}, \citenamefont {Ikenaga}, \citenamefont
  {Yamashita}, \citenamefont {Ueda},\ and\ \citenamefont
  {Kobayashi}}]{Alijani}%
  \BibitemOpen
  \bibfield  {author} {\bibinfo {author} {\bibfnamefont {V.}~\bibnamefont
  {Alijani}}, \bibinfo {author} {\bibfnamefont {S.}~\bibnamefont {Ouardi}},
  \bibinfo {author} {\bibfnamefont {G.~H.}\ \bibnamefont {Fecher}}, \bibinfo
  {author} {\bibfnamefont {J.}~\bibnamefont {Winterlik}}, \bibinfo {author}
  {\bibfnamefont {S.~S.}\ \bibnamefont {Naghavi}}, \bibinfo {author}
  {\bibfnamefont {X.}~\bibnamefont {Kozina}}, \bibinfo {author} {\bibfnamefont
  {G.}~\bibnamefont {Stryganyuk}}, \bibinfo {author} {\bibfnamefont
  {C.}~\bibnamefont {Felser}}, \bibinfo {author} {\bibfnamefont
  {E.}~\bibnamefont {Ikenaga}}, \bibinfo {author} {\bibfnamefont
  {Y.}~\bibnamefont {Yamashita}}, \bibinfo {author} {\bibfnamefont
  {S.}~\bibnamefont {Ueda}}, \ and\ \bibinfo {author} {\bibfnamefont
  {K.}~\bibnamefont {Kobayashi}},\ }\href {\doibase 10.1103/PhysRevB.84.224416}
  {\bibfield  {journal} {\bibinfo  {journal} {Phys. Rev. B}\ }\textbf {\bibinfo
  {volume} {84}},\ \bibinfo {pages} {224416} (\bibinfo {year}
  {2011})}\BibitemShut {NoStop}%
\bibitem [{\citenamefont {Felser}\ \emph {et~al.}(2007)\citenamefont {Felser},
  \citenamefont {Fecher},\ and\ \citenamefont {Balke}}]{Felser}%
  \BibitemOpen
  \bibfield  {author} {\bibinfo {author} {\bibfnamefont {C.}~\bibnamefont
  {Felser}}, \bibinfo {author} {\bibfnamefont {G.~H.}\ \bibnamefont {Fecher}},
  \ and\ \bibinfo {author} {\bibfnamefont {B.}~\bibnamefont {Balke}},\
  }\href@noop {} {\bibfield  {journal} {\bibinfo  {journal} {Angew. Chem. Int.
  Ed.}\ }\textbf {\bibinfo {volume} {46}},\ \bibinfo {pages} {668} (\bibinfo
  {year} {2007})}\BibitemShut {NoStop}%
\bibitem [{\citenamefont {Graf}\ \emph {et~al.}(2011)\citenamefont {Graf},
  \citenamefont {Felser},\ and\ \citenamefont {Parkin}}]{Graf1}%
  \BibitemOpen
  \bibfield  {author} {\bibinfo {author} {\bibfnamefont {T.}~\bibnamefont
  {Graf}}, \bibinfo {author} {\bibfnamefont {C.}~\bibnamefont {Felser}}, \ and\
  \bibinfo {author} {\bibfnamefont {S.~S.~P.}\ \bibnamefont {Parkin}},\
  }\href@noop {} {\bibfield  {journal} {\bibinfo  {journal} {Progress in Solid
  State Chemistry}\ }\textbf {\bibinfo {volume} {39}},\ \bibinfo {pages} {1}
  (\bibinfo {year} {2011})}\BibitemShut {NoStop}%
\bibitem [{\citenamefont {Graf}\ \emph {et~al.}(2013)\citenamefont {Graf},
  \citenamefont {Winterlik}, \citenamefont {Muchler}, \citenamefont {Fecher},
  \citenamefont {Felser},\ and\ \citenamefont {P.}}]{Graf2}%
  \BibitemOpen
  \bibfield  {author} {\bibinfo {author} {\bibfnamefont {T.}~\bibnamefont
  {Graf}}, \bibinfo {author} {\bibfnamefont {L.}~\bibnamefont {Winterlik}},
  \bibinfo {author} {\bibfnamefont {L.}~\bibnamefont {Muchler}}, \bibinfo
  {author} {\bibfnamefont {G.~H.}\ \bibnamefont {Fecher}}, \bibinfo {author}
  {\bibfnamefont {C.}~\bibnamefont {Felser}}, \ and\ \bibinfo {author}
  {\bibfnamefont {P.~S.~P.}\ \bibnamefont {P.}},\ }\href@noop {} {\bibfield
  {journal} {\bibinfo  {journal} {Handbook of Magnetic Materials}\ }\textbf
  {\bibinfo {volume} {21}},\ \bibinfo {pages} {1} (\bibinfo {year}
  {2013})}\BibitemShut {NoStop}%
\bibitem [{\citenamefont {Kreiner}\ \emph {et~al.}(2014)\citenamefont
  {Kreiner}, \citenamefont {Kalache}, \citenamefont {Hausdorf}, \citenamefont
  {Alijanin}, \citenamefont {Qian}, \citenamefont {Burkhardt}, \citenamefont
  {Ouardi},\ and\ \citenamefont {Felser}}]{Kreiner}%
  \BibitemOpen
  \bibfield  {author} {\bibinfo {author} {\bibfnamefont {G.}~\bibnamefont
  {Kreiner}}, \bibinfo {author} {\bibfnamefont {A.}~\bibnamefont {Kalache}},
  \bibinfo {author} {\bibfnamefont {S.}~\bibnamefont {Hausdorf}}, \bibinfo
  {author} {\bibfnamefont {V.}~\bibnamefont {Alijanin}}, \bibinfo {author}
  {\bibfnamefont {J.-F.}\ \bibnamefont {Qian}}, \bibinfo {author}
  {\bibfnamefont {U.}~\bibnamefont {Burkhardt}}, \bibinfo {author}
  {\bibfnamefont {S.}~\bibnamefont {Ouardi}}, \ and\ \bibinfo {author}
  {\bibfnamefont {C.}~\bibnamefont {Felser}},\ }\href@noop {} {\bibfield
  {journal} {\bibinfo  {journal} {Z. Anorg. Allg. Chem}\ }\textbf {\bibinfo
  {volume} {640}},\ \bibinfo {pages} {738} (\bibinfo {year}
  {2014})}\BibitemShut {NoStop}%
\bibitem [{\citenamefont {Roy}\ and\ \citenamefont
  {Chakrabarti}(2016)}]{roy16}%
  \BibitemOpen
  \bibfield  {author} {\bibinfo {author} {\bibfnamefont {T.}~\bibnamefont
  {Roy}}\ and\ \bibinfo {author} {\bibfnamefont {A.}~\bibnamefont
  {Chakrabarti}},\ }\href@noop {} {\bibfield  {journal} {\bibinfo  {journal}
  {arXiv:1603.08350}\ } (\bibinfo {year} {2016})}\BibitemShut {NoStop}%
\bibitem [{\citenamefont {Wollmann}\ \emph {et~al.}(2015)\citenamefont
  {Wollmann}, \citenamefont {Chadov}, \citenamefont {K\"ubler},\ and\
  \citenamefont {Felser}}]{woll15}%
  \BibitemOpen
  \bibfield  {author} {\bibinfo {author} {\bibfnamefont {L.}~\bibnamefont
  {Wollmann}}, \bibinfo {author} {\bibfnamefont {S.}~\bibnamefont {Chadov}},
  \bibinfo {author} {\bibfnamefont {J.}~\bibnamefont {K\"ubler}}, \ and\
  \bibinfo {author} {\bibfnamefont {C.}~\bibnamefont {Felser}},\ }\href
  {\doibase 10.1103/PhysRevB.92.064417} {\bibfield  {journal} {\bibinfo
  {journal} {Phys. Rev. B}\ }\textbf {\bibinfo {volume} {92}},\ \bibinfo
  {pages} {064417} (\bibinfo {year} {2015})}\BibitemShut {NoStop}%
\bibitem [{\citenamefont {Talapatra}\ \emph {et~al.}(2015)\citenamefont
  {Talapatra}, \citenamefont {Arr\'oyave}, \citenamefont {Entel}, \citenamefont
  {Valencia-Jaime},\ and\ \citenamefont {Romero}}]{tal15}%
  \BibitemOpen
  \bibfield  {author} {\bibinfo {author} {\bibfnamefont {A.}~\bibnamefont
  {Talapatra}}, \bibinfo {author} {\bibfnamefont {R.}~\bibnamefont
  {Arr\'oyave}}, \bibinfo {author} {\bibfnamefont {P.}~\bibnamefont {Entel}},
  \bibinfo {author} {\bibfnamefont {I.}~\bibnamefont {Valencia-Jaime}}, \ and\
  \bibinfo {author} {\bibfnamefont {A.~H.}\ \bibnamefont {Romero}},\ }\href
  {\doibase 10.1103/PhysRevB.92.054107} {\bibfield  {journal} {\bibinfo
  {journal} {Phys. Rev. B}\ }\textbf {\bibinfo {volume} {92}},\ \bibinfo
  {pages} {054107} (\bibinfo {year} {2015})}\BibitemShut {NoStop}%
\bibitem [{\citenamefont {Hongzhi}\ \emph {et~al.}(2013)\citenamefont
  {Hongzhi}, \citenamefont {Pengzhong}, \citenamefont {Guodong}, \citenamefont
  {Fanbin}, \citenamefont {Heyan}, \citenamefont {Enke}, \citenamefont
  {Wenhong},\ and\ \citenamefont {Guangheng}}]{hong13}%
  \BibitemOpen
  \bibfield  {author} {\bibinfo {author} {\bibfnamefont {L.}~\bibnamefont
  {Hongzhi}}, \bibinfo {author} {\bibfnamefont {J.}~\bibnamefont {Pengzhong}},
  \bibinfo {author} {\bibfnamefont {L.}~\bibnamefont {Guodong}}, \bibinfo
  {author} {\bibfnamefont {M.}~\bibnamefont {Fanbin}}, \bibinfo {author}
  {\bibfnamefont {L.}~\bibnamefont {Heyan}}, \bibinfo {author} {\bibfnamefont
  {L.}~\bibnamefont {Enke}}, \bibinfo {author} {\bibfnamefont {W.}~\bibnamefont
  {Wenhong}}, \ and\ \bibinfo {author} {\bibfnamefont {W.}~\bibnamefont
  {Guangheng}},\ }\href@noop {} {\bibfield  {journal} {\bibinfo  {journal}
  {Solid State Communications}\ }\textbf {\bibinfo {volume} {170}},\ \bibinfo
  {pages} {44–47} (\bibinfo {year} {2013})}\BibitemShut {NoStop}%
\bibitem [{\citenamefont {Winterlik}\ \emph
  {et~al.}(2012{\natexlab{b}})\citenamefont {Winterlik}, \citenamefont
  {Chadov}, \citenamefont {Gupta}, \citenamefont {Alijani}, \citenamefont
  {Gasi}, \citenamefont {Filsinger}, \citenamefont {Balke}, \citenamefont
  {Fecher}, \citenamefont {Jenkins}, \citenamefont {Casper}, \citenamefont
  {Kübler}, \citenamefont {Liu}, \citenamefont {Gao}, \citenamefont {Parkin},\
  and\ \citenamefont {Felser}}]{wint12}%
  \BibitemOpen
  \bibfield  {author} {\bibinfo {author} {\bibfnamefont {J.}~\bibnamefont
  {Winterlik}}, \bibinfo {author} {\bibfnamefont {S.}~\bibnamefont {Chadov}},
  \bibinfo {author} {\bibfnamefont {A.}~\bibnamefont {Gupta}}, \bibinfo
  {author} {\bibfnamefont {V.}~\bibnamefont {Alijani}}, \bibinfo {author}
  {\bibfnamefont {T.}~\bibnamefont {Gasi}}, \bibinfo {author} {\bibfnamefont
  {K.}~\bibnamefont {Filsinger}}, \bibinfo {author} {\bibfnamefont
  {B.}~\bibnamefont {Balke}}, \bibinfo {author} {\bibfnamefont {G.~H.}\
  \bibnamefont {Fecher}}, \bibinfo {author} {\bibfnamefont {C.~A.}\
  \bibnamefont {Jenkins}}, \bibinfo {author} {\bibfnamefont {F.}~\bibnamefont
  {Casper}}, \bibinfo {author} {\bibfnamefont {J.}~\bibnamefont {Kübler}},
  \bibinfo {author} {\bibfnamefont {G.-D.}\ \bibnamefont {Liu}}, \bibinfo
  {author} {\bibfnamefont {L.}~\bibnamefont {Gao}}, \bibinfo {author}
  {\bibfnamefont {S.~S.~P.}\ \bibnamefont {Parkin}}, \ and\ \bibinfo {author}
  {\bibfnamefont {C.}~\bibnamefont {Felser}},\ }\href {\doibase
  10.1002/adma.201201879} {\bibfield  {journal} {\bibinfo  {journal} {Advanced
  Materials}\ }\textbf {\bibinfo {volume} {24}},\ \bibinfo {pages} {6283}
  (\bibinfo {year} {2012}{\natexlab{b}})}\BibitemShut {NoStop}%
\bibitem [{\citenamefont {Wollmann}\ \emph {et~al.}(2014)\citenamefont
  {Wollmann}, \citenamefont {Chadov}, \citenamefont {K\"ubler},\ and\
  \citenamefont {Felser}}]{woll14}%
  \BibitemOpen
  \bibfield  {author} {\bibinfo {author} {\bibfnamefont {L.}~\bibnamefont
  {Wollmann}}, \bibinfo {author} {\bibfnamefont {S.}~\bibnamefont {Chadov}},
  \bibinfo {author} {\bibfnamefont {J.}~\bibnamefont {K\"ubler}}, \ and\
  \bibinfo {author} {\bibfnamefont {C.}~\bibnamefont {Felser}},\ }\href
  {\doibase 10.1103/PhysRevB.90.214420} {\bibfield  {journal} {\bibinfo
  {journal} {Phys. Rev. B}\ }\textbf {\bibinfo {volume} {90}},\ \bibinfo
  {pages} {214420} (\bibinfo {year} {2014})}\BibitemShut {NoStop}%
\bibitem [{\citenamefont {Kresse}\ and\ \citenamefont
  {Furthm\"{u}ller}(1996)}]{vasp}%
  \BibitemOpen
  \bibfield  {author} {\bibinfo {author} {\bibfnamefont {G.}~\bibnamefont
  {Kresse}}\ and\ \bibinfo {author} {\bibfnamefont {J.}~\bibnamefont
  {Furthm\"{u}ller}},\ }\href@noop {} {\bibfield  {journal} {\bibinfo
  {journal} {Comput. Mat. Sci.}\ }\textbf {\bibinfo {volume} {6}},\ \bibinfo
  {pages} {15} (\bibinfo {year} {1996})}\BibitemShut {NoStop}%
\bibitem [{\citenamefont {Bl\"{o}chl}(1994)}]{PAW}%
  \BibitemOpen
  \bibfield  {author} {\bibinfo {author} {\bibfnamefont {P.}~\bibnamefont
  {Bl\"{o}chl}},\ }\href@noop {} {\bibfield  {journal} {\bibinfo  {journal}
  {Phys. Rev. B}\ }\textbf {\bibinfo {volume} {50}},\ \bibinfo {pages} {17953}
  (\bibinfo {year} {1994})}\BibitemShut {NoStop}%
\bibitem [{\citenamefont {Perdew}\ \emph {et~al.}(1996)\citenamefont {Perdew},
  \citenamefont {Burke},\ and\ \citenamefont {Ernzerhof}}]{PBE}%
  \BibitemOpen
  \bibfield  {author} {\bibinfo {author} {\bibfnamefont {J.~P.}\ \bibnamefont
  {Perdew}}, \bibinfo {author} {\bibfnamefont {K.}~\bibnamefont {Burke}}, \
  and\ \bibinfo {author} {\bibfnamefont {M.}~\bibnamefont {Ernzerhof}},\ }\href
  {\doibase 10.1103/PhysRevLett.77.3865} {\bibfield  {journal} {\bibinfo
  {journal} {Phys. Rev. Lett.}\ }\textbf {\bibinfo {volume} {77}},\ \bibinfo
  {pages} {3865} (\bibinfo {year} {1996})}\BibitemShut {NoStop}%
\bibitem [{\citenamefont {Dewhurst}\ \emph {et~al.}(2016)\citenamefont
  {Dewhurst}, \citenamefont {Sharma},\ and\ \citenamefont {et~al.}}]{elk}%
  \BibitemOpen
  \bibfield  {author} {\bibinfo {author} {\bibfnamefont {K.}~\bibnamefont
  {Dewhurst}}, \bibinfo {author} {\bibfnamefont {S.}~\bibnamefont {Sharma}}, \
  and\ \bibinfo {author} {\bibnamefont {et~al.}},\ }\href@noop {} {\bibfield
  {journal} {\bibinfo  {journal} {http://elk.sourceforge.net/}\ } (\bibinfo
  {year} {2016})}\BibitemShut {NoStop}%
\bibitem [{\citenamefont {Gillessen}\ and\ \citenamefont
  {Dronskowski}(2010)}]{gill10}%
  \BibitemOpen
  \bibfield  {author} {\bibinfo {author} {\bibfnamefont {M.}~\bibnamefont
  {Gillessen}}\ and\ \bibinfo {author} {\bibfnamefont {R.}~\bibnamefont
  {Dronskowski}},\ }\href {\doibase 10.1002/jcc.21358} {\bibfield  {journal}
  {\bibinfo  {journal} {Journal of Computational Chemistry}\ }\textbf {\bibinfo
  {volume} {31}},\ \bibinfo {pages} {612} (\bibinfo {year} {2010})}\BibitemShut
  {NoStop}%
\bibitem [{\citenamefont {Gasi}\ \emph {et~al.}(2013)\citenamefont {Gasi},
  \citenamefont {Ksenofontov}, \citenamefont {Kiss}, \citenamefont {Chadov},
  \citenamefont {Nayak}, \citenamefont {Nicklas}, \citenamefont {Winterlik},
  \citenamefont {Schwall}, \citenamefont {Klaer}, \citenamefont {Adler},\ and\
  \citenamefont {Felser}}]{Gasi}%
  \BibitemOpen
  \bibfield  {author} {\bibinfo {author} {\bibfnamefont {T.}~\bibnamefont
  {Gasi}}, \bibinfo {author} {\bibfnamefont {V.}~\bibnamefont {Ksenofontov}},
  \bibinfo {author} {\bibfnamefont {J.}~\bibnamefont {Kiss}}, \bibinfo {author}
  {\bibfnamefont {S.}~\bibnamefont {Chadov}}, \bibinfo {author} {\bibfnamefont
  {A.~K.}\ \bibnamefont {Nayak}}, \bibinfo {author} {\bibfnamefont
  {M.}~\bibnamefont {Nicklas}}, \bibinfo {author} {\bibfnamefont
  {J.}~\bibnamefont {Winterlik}}, \bibinfo {author} {\bibfnamefont
  {M.}~\bibnamefont {Schwall}}, \bibinfo {author} {\bibfnamefont
  {P.}~\bibnamefont {Klaer}}, \bibinfo {author} {\bibfnamefont
  {P.}~\bibnamefont {Adler}}, \ and\ \bibinfo {author} {\bibfnamefont
  {C.}~\bibnamefont {Felser}},\ }\href {\doibase 10.1103/PhysRevB.87.064411}
  {\bibfield  {journal} {\bibinfo  {journal} {Phys. Rev. B}\ }\textbf {\bibinfo
  {volume} {87}},\ \bibinfo {pages} {064411} (\bibinfo {year}
  {2013})}\BibitemShut {NoStop}%
\bibitem [{\citenamefont {Felser}\ and\ \citenamefont {Hirohata}(2016)}]{book}%
  \BibitemOpen
  \bibfield  {author} {\bibinfo {author} {\bibfnamefont {C.}~\bibnamefont
  {Felser}}\ and\ \bibinfo {author} {\bibfnamefont {A.~E.}\ \bibnamefont
  {Hirohata}},\ }\href@noop {} {\bibfield  {journal} {\bibinfo  {journal}
  {Springer Series in Materials Science 222}\ } (\bibinfo {year}
  {2016})}\BibitemShut {NoStop}%
\bibitem [{\citenamefont {Suits}(1976)}]{Rh}%
  \BibitemOpen
  \bibfield  {author} {\bibinfo {author} {\bibfnamefont {J.}~\bibnamefont
  {Suits}},\ }\href@noop {} {\bibfield  {journal} {\bibinfo  {journal} {Solid
  State Commun.}\ }\textbf {\bibinfo {volume} {18}},\ \bibinfo {pages} {423}
  (\bibinfo {year} {1976})}\BibitemShut {NoStop}%
\bibitem [{\citenamefont {Barman}\ \emph {et~al.}(2007)\citenamefont {Barman},
  \citenamefont {Banik}, \citenamefont {Shukla}, \citenamefont {Kamal},\ and\
  \citenamefont {Chakrabarti}}]{bar07}%
  \BibitemOpen
  \bibfield  {author} {\bibinfo {author} {\bibfnamefont {S.~R.}\ \bibnamefont
  {Barman}}, \bibinfo {author} {\bibfnamefont {S.}~\bibnamefont {Banik}},
  \bibinfo {author} {\bibfnamefont {A.~K.}\ \bibnamefont {Shukla}}, \bibinfo
  {author} {\bibfnamefont {C.}~\bibnamefont {Kamal}}, \ and\ \bibinfo {author}
  {\bibfnamefont {A.}~\bibnamefont {Chakrabarti}},\ }\href
  {http://stacks.iop.org/0295-5075/80/i=5/a=57002} {\bibfield  {journal}
  {\bibinfo  {journal} {EPL (Europhysics Letters)}\ }\textbf {\bibinfo {volume}
  {80}},\ \bibinfo {pages} {57002} (\bibinfo {year} {2007})}\BibitemShut
  {NoStop}%
\bibitem [{\citenamefont {Zayak}\ \emph {et~al.}(2005)\citenamefont {Zayak},
  \citenamefont {Entel}, \citenamefont {Rabe}, \citenamefont {Adeagbo},\ and\
  \citenamefont {Acet}}]{zay05}%
  \BibitemOpen
  \bibfield  {author} {\bibinfo {author} {\bibfnamefont {A.~T.}\ \bibnamefont
  {Zayak}}, \bibinfo {author} {\bibfnamefont {P.}~\bibnamefont {Entel}},
  \bibinfo {author} {\bibfnamefont {K.~M.}\ \bibnamefont {Rabe}}, \bibinfo
  {author} {\bibfnamefont {W.~A.}\ \bibnamefont {Adeagbo}}, \ and\ \bibinfo
  {author} {\bibfnamefont {M.}~\bibnamefont {Acet}},\ }\href {\doibase
  10.1103/PhysRevB.72.054113} {\bibfield  {journal} {\bibinfo  {journal} {Phys.
  Rev. B}\ }\textbf {\bibinfo {volume} {72}},\ \bibinfo {pages} {054113}
  (\bibinfo {year} {2005})}\BibitemShut {NoStop}%
\bibitem [{\citenamefont {Paul}\ \emph {et~al.}(2015)\citenamefont {Paul},
  \citenamefont {Sanyal},\ and\ \citenamefont {Ghosh}}]{paul15}%
  \BibitemOpen
  \bibfield  {author} {\bibinfo {author} {\bibfnamefont {S.}~\bibnamefont
  {Paul}}, \bibinfo {author} {\bibfnamefont {B.}~\bibnamefont {Sanyal}}, \ and\
  \bibinfo {author} {\bibfnamefont {S.}~\bibnamefont {Ghosh}},\ }\href
  {http://stacks.iop.org/0953-8984/27/i=3/a=035401} {\bibfield  {journal}
  {\bibinfo  {journal} {Journal of Physics: Condensed Matter}\ }\textbf
  {\bibinfo {volume} {27}},\ \bibinfo {pages} {035401} (\bibinfo {year}
  {2015})}\BibitemShut {NoStop}%
\bibitem [{\citenamefont {Ivanov}\ \emph {et~al.}(1973)\citenamefont {Ivanov},
  \citenamefont {Solina}, \citenamefont {Demshira},\ and\ \citenamefont
  {Magat}}]{FePt_expt}%
  \BibitemOpen
  \bibfield  {author} {\bibinfo {author} {\bibfnamefont {O.~A.}\ \bibnamefont
  {Ivanov}}, \bibinfo {author} {\bibfnamefont {L.~V.}\ \bibnamefont {Solina}},
  \bibinfo {author} {\bibfnamefont {V.~A.}\ \bibnamefont {Demshira}}, \ and\
  \bibinfo {author} {\bibfnamefont {L.~M.}\ \bibnamefont {Magat}},\ }\href@noop
  {} {\bibfield  {journal} {\bibinfo  {journal} {Phys. Met. Metallov.}\
  }\textbf {\bibinfo {volume} {35}},\ \bibinfo {pages} {92} (\bibinfo {year}
  {1973})}\BibitemShut {NoStop}%
\bibitem [{\citenamefont {Fichtner}\ \emph {et~al.}(2015)\citenamefont
  {Fichtner}, \citenamefont {Wang}, \citenamefont {Levin}, \citenamefont
  {Kreiner}, \citenamefont {Mejia}, \citenamefont {Fabbrici}, \citenamefont
  {Albertini},\ and\ \citenamefont {Felser}}]{Co2NiGa}%
  \BibitemOpen
  \bibfield  {author} {\bibinfo {author} {\bibfnamefont {T.}~\bibnamefont
  {Fichtner}}, \bibinfo {author} {\bibfnamefont {C.}~\bibnamefont {Wang}},
  \bibinfo {author} {\bibfnamefont {A.}~\bibnamefont {Levin}}, \bibinfo
  {author} {\bibfnamefont {G.}~\bibnamefont {Kreiner}}, \bibinfo {author}
  {\bibfnamefont {C.}~\bibnamefont {Mejia}}, \bibinfo {author} {\bibfnamefont
  {S.}~\bibnamefont {Fabbrici}}, \bibinfo {author} {\bibfnamefont
  {F.}~\bibnamefont {Albertini}}, \ and\ \bibinfo {author} {\bibfnamefont
  {C.}~\bibnamefont {Felser}},\ }\href@noop {} {\bibfield  {journal} {\bibinfo
  {journal} {Metals}\ }\textbf {\bibinfo {volume} {5}},\ \bibinfo {pages} {484}
  (\bibinfo {year} {2015})}\BibitemShut {NoStop}%
\bibitem [{\citenamefont {Gabor}\ \emph {et~al.}(2011)\citenamefont {Gabor},
  \citenamefont {Petrisor}, \citenamefont {Tiusan}, \citenamefont {Hehn},\ and\
  \citenamefont {Petrisor}}]{Co2FeAl1}%
  \BibitemOpen
  \bibfield  {author} {\bibinfo {author} {\bibfnamefont {M.~S.}\ \bibnamefont
  {Gabor}}, \bibinfo {author} {\bibfnamefont {T.}~\bibnamefont {Petrisor}},
  \bibinfo {author} {\bibfnamefont {C.}~\bibnamefont {Tiusan}}, \bibinfo
  {author} {\bibfnamefont {M.}~\bibnamefont {Hehn}}, \ and\ \bibinfo {author}
  {\bibfnamefont {T.}~\bibnamefont {Petrisor}},\ }\href {\doibase
  10.1103/PhysRevB.84.134413} {\bibfield  {journal} {\bibinfo  {journal} {Phys.
  Rev. B}\ }\textbf {\bibinfo {volume} {84}},\ \bibinfo {pages} {134413}
  (\bibinfo {year} {2011})}\BibitemShut {NoStop}%
\bibitem [{\citenamefont {Husain}\ \emph {et~al.}(2016)\citenamefont {Husain},
  \citenamefont {Akansel}, \citenamefont {Svedlindh},\ and\ \citenamefont
  {Chaudhary}}]{Co2FeAl2}%
  \BibitemOpen
  \bibfield  {author} {\bibinfo {author} {\bibfnamefont {S.}~\bibnamefont
  {Husain}}, \bibinfo {author} {\bibfnamefont {A.}~\bibnamefont {Akansel},
  \bibfnamefont {S.and~Kumar}}, \bibinfo {author} {\bibfnamefont
  {P.}~\bibnamefont {Svedlindh}}, \ and\ \bibinfo {author} {\bibfnamefont
  {S.}~\bibnamefont {Chaudhary}},\ }\href@noop {} {\bibfield  {journal}
  {\bibinfo  {journal} {Scientific Reports}\ }\textbf {\bibinfo {volume}
  {432}},\ \bibinfo {pages} {28692} (\bibinfo {year} {2016})}\BibitemShut
  {NoStop}%
\bibitem [{\citenamefont {Zhang}\ \emph {et~al.}(2004)\citenamefont {Zhang},
  \citenamefont {Brück}, \citenamefont {de~Boer}, \citenamefont {Li},\ and\
  \citenamefont {Wu}}]{Co2FeGa}%
  \BibitemOpen
  \bibfield  {author} {\bibinfo {author} {\bibfnamefont {M.}~\bibnamefont
  {Zhang}}, \bibinfo {author} {\bibfnamefont {E.}~\bibnamefont {Brück}},
  \bibinfo {author} {\bibfnamefont {F.~R.}\ \bibnamefont {de~Boer}}, \bibinfo
  {author} {\bibfnamefont {Z.}~\bibnamefont {Li}}, \ and\ \bibinfo {author}
  {\bibfnamefont {G.}~\bibnamefont {Wu}},\ }\href
  {http://stacks.iop.org/0022-3727/37/i=15/a=001} {\bibfield  {journal}
  {\bibinfo  {journal} {Journal of Physics D: Applied Physics}\ }\textbf
  {\bibinfo {volume} {37}},\ \bibinfo {pages} {2049} (\bibinfo {year}
  {2004})}\BibitemShut {NoStop}%
\bibitem [{\citenamefont {Uvarov1}\ \emph {et~al.}(2012)\citenamefont
  {Uvarov1}, \citenamefont {Kudryavtsev1}, \citenamefont {Kravets},
  \citenamefont {Vovk}, \citenamefont {Borges}, \citenamefont {Godinho},\ and\
  \citenamefont {Korenivski}}]{Co2FeGe}%
  \BibitemOpen
  \bibfield  {author} {\bibinfo {author} {\bibfnamefont {N.}~\bibnamefont
  {Uvarov1}}, \bibinfo {author} {\bibfnamefont {Y.}~\bibnamefont
  {Kudryavtsev1}}, \bibinfo {author} {\bibfnamefont {A.}~\bibnamefont
  {Kravets}}, \bibinfo {author} {\bibfnamefont {Y.}~\bibnamefont {Vovk}},
  \bibinfo {author} {\bibfnamefont {R.}~\bibnamefont {Borges}}, \bibinfo
  {author} {\bibfnamefont {M.}~\bibnamefont {Godinho}}, \ and\ \bibinfo
  {author} {\bibfnamefont {V.}~\bibnamefont {Korenivski}},\ }\href@noop {}
  {\bibfield  {journal} {\bibinfo  {journal} {J. Appl. Phys.}\ }\textbf
  {\bibinfo {volume} {112}},\ \bibinfo {pages} {063909} (\bibinfo {year}
  {2012})}\BibitemShut {NoStop}%
\bibitem [{\citenamefont {Zhang}\ \emph {et~al.}(2013)\citenamefont {Zhang},
  \citenamefont {Wang}, \citenamefont {Zhang}, \citenamefont {Liu},
  \citenamefont {Ma},\ and\ \citenamefont {H.}}]{Zhang}%
  \BibitemOpen
  \bibfield  {author} {\bibinfo {author} {\bibfnamefont {Y.~J.}\ \bibnamefont
  {Zhang}}, \bibinfo {author} {\bibfnamefont {W.~H.}\ \bibnamefont {Wang}},
  \bibinfo {author} {\bibfnamefont {H.~G.}\ \bibnamefont {Zhang}}, \bibinfo
  {author} {\bibfnamefont {E.~K.}\ \bibnamefont {Liu}}, \bibinfo {author}
  {\bibfnamefont {R.~S.}\ \bibnamefont {Ma}}, \ and\ \bibinfo {author}
  {\bibfnamefont {W.~G.}\ \bibnamefont {H.}},\ }\href@noop {} {\bibfield
  {journal} {\bibinfo  {journal} {Physica B: Condensed Matter}\ }\textbf
  {\bibinfo {volume} {87}},\ \bibinfo {pages} {86} (\bibinfo {year}
  {2013})}\BibitemShut {NoStop}%
\bibitem [{\citenamefont {Buschow}\ \emph {et~al.}(1983)\citenamefont
  {Buschow}, \citenamefont {van Engen},\ and\ \citenamefont
  {Jongebreur}}]{Fe2NiAl2}%
  \BibitemOpen
  \bibfield  {author} {\bibinfo {author} {\bibfnamefont {K.}~\bibnamefont
  {Buschow}}, \bibinfo {author} {\bibfnamefont {P.}~\bibnamefont {van Engen}},
  \ and\ \bibinfo {author} {\bibfnamefont {R.}~\bibnamefont {Jongebreur}},\
  }\href@noop {} {\bibfield  {journal} {\bibinfo  {journal} {J. Magn. Magn.
  Mater.}\ }\textbf {\bibinfo {volume} {38}},\ \bibinfo {pages} {1} (\bibinfo
  {year} {1983})}\BibitemShut {NoStop}%
\bibitem [{\citenamefont {Csanad}\ \emph {et~al.}(2004)\citenamefont {Csanad},
  \citenamefont {Csorgo},\ and\ \citenamefont {Lorstad}}]{Fe2NiAl3}%
  \BibitemOpen
  \bibfield  {author} {\bibinfo {author} {\bibfnamefont {M.}~\bibnamefont
  {Csanad}}, \bibinfo {author} {\bibfnamefont {T.}~\bibnamefont {Csorgo}}, \
  and\ \bibinfo {author} {\bibfnamefont {B.}~\bibnamefont {Lorstad}},\
  }\href@noop {} {\bibfield  {journal} {\bibinfo  {journal} {Nukleonika}\
  }\textbf {\bibinfo {volume} {49}},\ \bibinfo {pages} {S49} (\bibinfo {year}
  {2004})}\BibitemShut {NoStop}%
\bibitem [{\citenamefont {Ren}\ \emph {et~al.}(2010)\citenamefont {Ren},
  \citenamefont {Li},\ and\ \citenamefont {Luo}}]{Fe2NiGe}%
  \BibitemOpen
  \bibfield  {author} {\bibinfo {author} {\bibfnamefont {Z.}~\bibnamefont
  {Ren}}, \bibinfo {author} {\bibfnamefont {S.~T.}\ \bibnamefont {Li}}, \ and\
  \bibinfo {author} {\bibfnamefont {H.~Z.}\ \bibnamefont {Luo}},\ }\href@noop
  {} {\bibfield  {journal} {\bibinfo  {journal} {Physica B: Condensed Matter}\
  }\textbf {\bibinfo {volume} {405}},\ \bibinfo {pages} {2840} (\bibinfo {year}
  {2010})}\BibitemShut {NoStop}%
\end{thebibliography}
%
\end{document}